# Magnetoelectric properties of 500 nm $Cr_2O_3$ films

P. Borisov,[1a] T. Ashida,[2] T. Nozaki,[2] M. Sahashi,[2] D. Lederman[1,3]

[1] Department of Physics and Astronomy, West Virginia University, Morgantown WV 26506-6315, USA

[2] Department of Electronic Engineering, Tohoku University, Sendai, 980-0845, Japan

[3] Department of Physics, University of California, Santa Cruz, CA 95064, USA

**Abstract**

The linear magnetoelectric effect was measured in 500 nm $Cr_2O_3$ films grown by rf sputtering on $Al_2O_3$ substrates between top and bottom thin film Pt electrodes. Magnetoelectric susceptibility was measured directly by applying an AC electric field and measuring the induced AC magnetic moment using superconducting quantum interference device magnetometry. A linear dependence of the induced AC magnetic moment on the AC electric field amplitude was found. The temperature dependence of the magnetoelectric susceptibility agreed qualitatively and quantitatively with prior measurements of bulk single crystals, but the characteristic temperatures of the film were lower than those of single crystals. It was also possible to reverse the sign of the magnetoelectric susceptibility by reversing the sign of the magnetic field applied during cooling through the Néel temperature. A competition between total magnetoelectric and Zeeman energies is proposed to explain the difference between film and bulk $Cr_2O_3$ regarding the cooling field dependence of the magnetoelectric effect.

---

[a] Email of corresponding author: pavel.borisov@mail.wvu.edu

**Introduction**

Chromium oxide $Cr_2O_3$ has been intensively studied during the last 10 years because it is possible to modulate the exchange bias effect in heterostructures using electric fields [1 - 16]. In the antiferromagnetic (AF) state of $Cr_2O_3$, an applied electric field induces the magnetization, while an external magnetic field induces the dielectric polarization. The same magnetic symmetry that allows the linear magnetoelectric (ME) effect in $Cr_2O_3$ enables the existence of the uncompensated magnetization at the $Cr_2O_3$ (0001) surface which is insensitive to the roughness [5, 17]. The ME effect can be used to modify the AF domain structure and to switch between two types of 180° AF domains [18]. By modulating the interfacial magnetic moment in $Cr_2O_3$, it is possible to control the magnetization in an adjacent exchange coupled ferromagnetic (FM) layer [1, 3]. That is, a spin valve device can be built [2, 4] that allows for magnetic bit writing via the applied electric field.

The relatively small ME response of $Cr_2O_3$ makes its detection a complicated task, particularly in thin films, hence $Cr_2O_3$ can also be considered as a prototype material to test experimental approaches and techniques for measurements of the ME effect in thin film systems.

The ME effect in $Cr_2O_3$ has been measured previously in bulk single crystal samples [19,20]. Prior to the measurements, a single domain state was created via so-called ME field cooling (FC), where both electric ($E_{fc}$) and magnetic ($H_{fc}$) fields were applied by FC the sample from a temperature above the Neel point, $T_N$=308.5K. Two types of 180° AF domains can be formed in $Cr_2O_3$, with the spins aligned along [0001] $Cr_2O_3$. The ME susceptibility of the domains is defined as $\alpha=\mu_0 M/E$, where $M$ and $E$ denote the induced magnetization and the applied electric field, respectively, and $\mu_0$ is the permeability of free space. In $Cr_2O_3$, the ME susceptibilities of the two domains have the same magnitude but opposite signs. Therefore, during the ME FC procedure the two domains have increased or lowered total ME energy $\alpha E_{fc}H_{fc}$, depending on the sign of $\alpha$, so only one domain type is formed if a large enough $E_{fc}H_{fc}$ product is realized. In particular, the sign of the single domain ME susceptibility $\alpha$ can be reversed if two subsequent ME FC procedures are performed in the same magnetic FC field $\boldsymbol{H}_{fc}$, but in oppositely directed electric FC fields $\pm\boldsymbol{E}_{fc}$.

In this paper we show that the ME response of 500 nm films of $Cr_2O_3$ correlates with the boundary magnetization in $Cr_2O_3$ and that the sign of the ME susceptibility can be switched completely via cooling the sample in magnetic fields of opposite directions, while maintaining the electric zero field, $\boldsymbol{E}_{fc}=0$. This ability to switch the sign of $\alpha$ is a consequence of the reduced $Cr_2O_3$ thickness where the total Zeeman energy associated with the surface magnetization in $Cr_2O_3$ becomes comparable to the total ME energy associated with the crystal volume. This conclusion explains previous results [8,11 -13] of the electric switching of the exchange bias field in all-thin film systems (0001) $Cr_2O_3$/[Co/Pt]$_n$, where it was found that much higher field products $E_{fc}H_{fc}$ are required to achieve the switching compared to bulk single crystals of $Cr_2O_3$.

**Experiment**

The $Cr_2O_3$ films were grown by reactive rf sputtering on (0001) $Al_2O_3$ substrates using a metal Cr target in an Ar + $O_2$ atmosphere at a substrate temperature of 773 K (base pressure < 1 × 10$^{-6}$ Pa). Bottom and top Pt electrodes with thicknesses of 25 nm were sputtered on $Al_2O_3$ substrates and $Cr_2O_3$ films, respectively, using shadow masking (see inset to Fig. 3). ME susceptibilities were calculated using the following sample dimensions: electrode surface areas of 10.3 mm² and 9.5 mm² (relative error of ~15%) and sample thicknesses of 0.5 mm and 500 nm for $Cr_2O_3$ single crystal and film samples, respectively. Electrical contacts were made using copper wires and silver paint.

Structural properties were studied by X-ray diffraction (XRD). The XRD scans were performed in two modes: standard $\omega$-$2\theta$ Bragg geometry (out-of-plane scan) using D8 Advance (Bruker AXS) and horizontal (in-plane) $\phi$-$2\theta_\chi$ geometry using the ATX-G setup from Rigaku. Cu $K_\alpha$ X-ray sources were used in both cases. In addition, cross-sectional transmission electron microscopy (TEM) was performed using the H-9000NAR setup from Hitachi High Technologies.

The total thickness of the films has been determined using a probe-type step profiler and double-checked by cross-sectional TEM on a 250nm $Cr_2O_3$ film. We found a maximal 5% deviation of the total film thickness value from the nominal one.

A commercial magnetometer setup (MPMS from Quantum Design) equipped with a superconducting quantum interference device (SQUID) sensor was used for magnetic and ME measurements. The latter were performed by measuring the AC magnetic moment amplitudes $m'$ and $m''$, $m=m'\cos(2\pi ft)+m''\sin(2\pi ft)$, excited by an applied AC electric field, $E=E_{max}\cos(2\pi ft)$, where the frequency $f$ was chosen as $f$=17 Hz. For more details about this technique see [21]. The ME response was associated with $m'$ only, while $m''$=0 was within error bars, as expected for low frequency ME measurements in bulk $Cr_2O_3$ [21]. Simultaneous application of magnetic (up to 70 kOe) and electric (up to 200 V) DC fields was also possible. Electrical contacts were chosen such that positive electric and magnetic fields pointed in the same direction as positive magnetic moments. The first $E$-field half-period cycle was positive in the time dependence.

Magnetic measurements were performed using a standard sample holder. Thermoremanent magnetization (TRM) was measured under zero-field conditions upon warming the sample from 5 K to 320 K after FC from 320 K to 5 K in a magnetic field applied along the [0001] $Cr_2O_3$, that is, out of plane and along the spin easy axis. Averaged $Al_2O_3$ substrate-related magnetic remanence was calculated for temperatures >310K (maximum of ~$2\times10^{-7}$ emu) and then subtracted from experimental data to make the $Cr_2O_3$ related phase transitions more visible.

**Results and discussion**

Figure 1 (a) shows the XRD $\omega$-$2\theta$ (out of plane) scan for the (0001) $Al_2O_3$/ Pt 25 / $Cr_2O_3$ 500 (nm) sample. A comparison with the same scan on a (0001) $Al_2O_3$ / Pt 25 (nm) sample without the $Cr_2O_3$ layer demonstrates that the (111) and (222) reflections from (111) Pt bottom electrode overlap with the (0006) and (000$\underline{12}$) reflections from (0001) $Cr_2O_3$. The expected film thickness effect on the XRD peak intensities is mostly compensated by the weaker coherent X-ray scattering amplitudes of (0001) $Cr_2O_3$ reflections with respect to (111) Pt ones. No other oxide phases (for example $CrO_2$ or $CrO_3$) or alternative $Cr_2O_3$ orientations were observed. In order to verify the in-plane and out-of-plane orientation of the $Cr_2O_3$ film, horizontal XRD scans were performed with the scattering vector $q$ corresponding to the $Al_2O_3$ (10-10) [green line] and (11-20) [red line] (Fig. 1b) $Al_2O_3$ substrate planes on the (0001) $Al_2O_3$/ Pt 25 / $Cr_2O_3$ 500 (nm) sample. The results confirm the (0001) orientation of the $Cr_2O_3$ layer, with the corresponding in plane orientation $Al_2O_3$ [10-10] || $Cr_2O_3$ [10-10], though we observed a weak $Cr_2O_3$ peak for $Al_2O_3$ [11-20] || $Cr_2O_3$ [10-10], i.e. $Cr_2O_3$ rotated in-plane by 30° with respect to $Al_2O_3$. The latter alternative in-plane orientation was more pronounced in thinner (20nm) $Cr_2O_3$ films grown on (0001) $Al_2O_3$ / (111) Pt [16], hence, this feature seems to be related to crystallographic defects due to the interface (111) Pt / (0001) $Cr_2O_3$.

Since no reflections from Pt bottom electrodes were observed in $\phi$-$2\theta_\chi$ (in-plane) scans, due to the small probing depth and relatively thick $Cr_2O_3$ layer, we repeated the same scans for (0001) $Al_2O_3$ / Pt 25 (nm) samples and found Pt {2-20} peaks (not shown) for both cases of $q$ probing crystal planes parallel to $Al_2O_3$ (10-10) or $Al_2O_3$ (11-20).

The cross-sectional TEM image shown in Fig. 2 confirms crystalline (0001) $Cr_2O_3$ quality, while the Pt bottom electrode shows different in-plane structural domains. In conclusion, the structural studies verified single crystalline quality of the $Cr_2O_3$ films. Similar quality films have been used to demonstrate successful electric switching of the exchange bias field associated with the heterostructures $Al_2O_3$ (0001) / Pt / $Cr_2O_3$ (0001) / Pt / Co / Pt [8, 13].

Figure 3 shows AC magnetic moment amplitude $m'$ vs. $E$-field amplitude, $E_{max}$, measured at 250 K. In order to prepare a single AF domain state, the sample was cooled down in $H_{fc}$=10 kOe, as discussed below. A linear dependence was found, confirming that the ME effect in 500 nm $Cr_2O_3$ is qualitatively similar to the one in the bulk $Cr_2O_3$. Here, $m'=(V/\mu_0)\alpha E_{max}$, where $V$ is the $Cr_2O_3$ volume. The absolute value of the corresponding ME susceptibility was calculated from the slope, $\alpha = 4.6 \pm 0.3$ ps/m. ME measurements on another $Cr_2O_3$ film sample grown and field cooled under identical conditions yielded $\alpha = 4.1 \pm 0.3$ ps/m, that is, our results were reproducible between the films from the same batch. Also, the absolute values of the ME susceptibility for the film samples are in agreement with the single crystal data as discussed below.

The temperature dependence of the ME susceptibility $\alpha=\mu_0 m'/(V E_{max})$ is shown in Fig. 4. The film was cooled from 320 K to 100 K in magnetic fields 1 kOe or 10 kOe, and then measured while warming from 5 K. $E_{max}$ values were adjusted in different temperature ranges in order to achieve a compromise between a low leakage current and a large signal magnitude. Note the thermal activation character of the electrical conductivity in $Cr_2O_3$ [22]. Typical AC field amplitude values were 62.4 kV/cm between 5 K and 95 K (maximal leakage currents $i_{max} < 10$ nA), 31.2 kV/cm between 97 K and 150 K ($i_{max} < 10$ nA), and 6.24 kV/cm ($i_{max} < 300$ nA) or 20 kV/cm ($i_{max} < 3$ μA) at temperatures above 150 K. The largest measurement temperature was limited to 310 K in order to minimize the risk of a dielectric breakdown in the film.

Figure 4a shows a comparison of the ME susceptibilities of the film and bulk single crystal samples. The error bars refer to statistical uncertainties determined from the standard error of the average of multiple measurements. The single crystal was cooled in $H_{fc}$=30 kOe and $E_{fc}$=1 kV/cm. Prior to the measurements, the film sample was cooled in a magnetic field only. The largest ME response was found after FC in $H_{fc}$ ≥10kOe, with the magnitude of ME susceptibility $\alpha$ being comparable to the bulk single crystal values. The film cooled in a smaller $H_{fc}$ = 1 kOe showed a decrease in $|\alpha|$ of 30%, most likely because the cooling field was not strong enough to align the complete $Cr_2O_3$ surface in one single direction. This caused the two 180° AF domains to form with opposite signs of $\alpha$. This explanation assumes that the relative distribution of the AF domains is controlled by the boundary magnetization distribution. The results of DC magnetic measurements discussed below support this hypothesis.

In the film sample, ME susceptibility $\alpha$ shows a sign reversal at low temperatures at ~70 K (Fig. 4b) and a maximum below room temperature at ~250 K (Fig. 4a), with the signal disappearing above ~300 K, that is, above $T_N$ (Fig. 4c, film data were measured with different AC amplitudes, 6.24 kV/cm or 20.0 kV/cm, hence different relative errors in $\alpha$). These results are similar to what was observed in the bulk crystal sample (Fig. 4). As discussed below, we were also able to identify $T_N$ ~ 300 K from DC magnetic measurements on films (as will be shown in Fig. 6). Due to the difference in $T_N$ values between the film and the bulk single crystal samples, the relevant temperatures are also lower in the case of the films, cf. the response at 80K, 260K, and 310K in the case of a single crystal. A relatively unlikely scenario of the weak DC moment in $Cr_2O_3$ being somehow coupled to the ME AC susceptibility response can be excluded, since the sign reversal was found at ~70 K in the ME response (Fig. 4b) but was not observed in the DC magnetization measurements (Fig. 6).

The magnetic field applied during FC played the predominant role in the outcome of the ME measurements in films. Cooling in $H_{fc}$ = 0 resulted in the absence of a ME response (Fig. 5a), while subsequent cooling in two fields of the same magnitude but applied in opposite directions switched the sign of $\alpha$ (Fig. 5b). That is, the boundary magnetization couples directly to the external magnetic field and allows AF domain reversal [9].

We attempted to achieve true ME switching of $\alpha$ by applying the same $H_{fc}$ and reversing the direction of $E_{fc}$ applied during ME cooling from 310 K, but no switching was found using this procedure. The largest field product of $|H_{fc}|$ =50 kOe and $|E_{fc}|$ = 60 kV/cm was not successful, while the films were destroyed for $E_{fc} \geq 80$ kV/cm due to a dielectric breakdown. The lower breakdown field value when compared to previous reports [8, 11 - 13] is likely due to the larger electrode size in our case.

The results of the dc magnetic measurements are shown in Fig. 6. The remanent magnetic moment peaks at ~200 K and disappears above $T_C$ ~300 K, which agrees well with the critical temperature identified from the ME measurements (Fig. 4c), that is, with $T_N$. The temperature dependence is similar to prior results from thinner (26 – 60 nm) $Cr_2O_3$ films [9, 14]. In thinner films, it has been demonstrated that the uncompensated magnetic moment originates from the boundary magnetization coupled to the AF order parameter and is insensitive to the surface roughness [5, 17, 23]. The largest magnetic remanence occurred in the temperature interval between 195 K and 205 K. The magnetic moment was saturated for $H_{fc} \geq$ 2.5 kOe (Fig. 6b). Prior papers by Fallarino *et al.* [9, 14] considered the isothermal switching of the AF domain state in $Cr_2O_3$ films at $T < T_N$ in $H_{fc} \geq 10$ kOe, and found that the corresponding temperature at which the isothermal switching occurs, $T_f$, scales as $T_f = T_N - H_{fc}^2/b$, i.e. $T_f < T_N$. By using the parameter $b$ from [14] for 60 nm $Cr_2O_3$, as well as $T_N = 300$ K and $H_{fc} = 2.5$ kOe, we obtain $T_f = 299.99$K $\approx T_N$, that is, the Zeeman energy-induced AF state reversal via FC in $H_{fc} = 2.5$ kOe is likely not distinguishable from an isothermal effect.

As in Ref. 5, we recalculate the absolute value of the uncompensated moment associated with the boundary magnetization in $Cr_2O_3$ films (Fig. 6) into moment per hexagonal unit cell area, 64 Å$^2$, of the (0001) $Cr_2O_3$ film surface (total area 16.3

mm$^2$). Using the maximum observed saturation moment of ~$1.1\times10^{-6}$ emu, we obtain 4.7 $\mu_B$ per unit cell area, which is close to the 5.6 $\mu_B$ obtained for prior measurements of a 200 nm Cr$_2$O$_3$ film and explained by two Cr$^{3+}$ spin moments of 3$\mu_B$ (theoretical value) located in each hexagonal unit area [5]. However, this conclusion depends on the (0001) surface termination and spin structure. Assuming the (0001) Cr$_2$O$_3$ surface terminates as a half-cut through the Cr$^{3+}$ buckled layer between two adjacent closed-packed oxygen layers, as shown in Fig. 7 and in agreement with the literature [24 - 26], we calculate one Cr$^{3+}$ spin per hexagonal unit area of a single surface, or two spins in total, if top and bottom surfaces are considered. Alternatively, one can also consider a complete oxygen layer at the bottom surface and a complete buckled Cr$^{3+}$ layer located on the top surface above the oxygen layer. In both cases, some surface spins would have to be oriented along $H_{fc}$, that is, reversed with respect to the bulk structure, in order to yield a non-zero magnetic moment, in total. The representation of the magnetic moments shown in Fig. 7 is an idealization. Surface magnetization in (0001) Cr$_2$O$_3$ has been verified both experimentally and theoretically in Ref. 23, but the exact relationship between surface and bulk spins is still unclear. The observed 19% discrepancy between our result, 4.7$\mu_B$ per hexagonal unit cell area, and the moment found in Ref. 5, 5.7$\mu_B$, could be explained by the 15% relative error in our electrode area estimation.

Prior studies [18 - 20] of the ME properties of Cr$_2$O$_3$ single crystals reported that the 180° AF domain type can be selected by applying $E_{fc}$ and $H_{fc}$ parallel or antiparallel to each other and to the [0001] *c*-axis during ME FC. In the discussion below we assume that the magnetic field direction is always positive and applied along the *c*-axis. However, the same literature reported that $H_{fc}$ applied by itself during FC had some effect on the sign of the resulting ME susceptibility [18 -20]. The origin of the effect was not clear, since both types of AF domains should have the same magnetic susceptibility. Astrov [19] found that $H_{fc}$ = 500 Oe was enough to reverse the sign of the ME susceptibility in single crystals. A weak negative ME response at *T* = 150 K was obtained for single crystals (0001) Cr$_2$O$_3$ in Ref. 3 after a purely magnetic FC in a positive $H_{fc}$ = 5 kOe. Analysis of the spin structures in $Cr_2O_3$ using spherical neutron polarimetry [27 - 29] demonstrated that ME cooling in antiparallel $E_{fc}$ and $H_{fc}$ yields the AF domain type with the positive top magnetization, as shown in Fig. 7. We propose to call this a B type AF domain, distinguished from an A type domain by 180° rotation of all spins. The surface magnetization of a $Cr_2O_3$ single crystal prepared by ME cooling in parallel $E_{fc}$ and $H_{fc}$ was analyzed in Ref. 5 using spin-polarized photoemission spectroscopy results, and an opposite, that is, negative surface magnetization was obtained for this type A domain. Furthermore, by analyzing the ME results from the literature [3, 30, 31] we assign a positive and negative $\alpha$ for the temperature region > 80 K to domain types A and B, respectively. That is, our experimental results of $\alpha < 0$ at $T > 80$ K after FC in positive magnetic fields can be explained by the formation of the type B domain with the positive top boundary magnetization aligned parallel to $H_{fc}$.

The failure to achieve an AF single domain reversal via the ME FC procedure can be explained by the fact that with the reduced thickness of $Cr_2O_3$ single crystal film, the Zeeman energy, $\mu_0 m_S H_{fc}$, associated with the surface magnetic moment $m_S$, becomes comparable to the total ME energy, $\alpha E_{fc} H_{fc} V$, associated with the total volume $V$. Thus, purely magnetic switching via $\mu_0 m_S H_{fc}$ competes with ME switching via the $\alpha E_{fc} H_{fc} V$ energy term, when the corresponding energy terms promote the growth of the single domain type B and A, respectively. In order to estimate the corresponding energies of our samples, $T = 297.5$ K was chosen as the relevant temperature being relatively close to $T_N$, but at the same time, where the corresponding ME susceptibility was measured with sufficient precision, $\alpha = 1.7 \pm 0.2$ ps/m, after FC in 10 kOe (Fig. 5a). The corresponding DC moment at $T = 297.5$ K was obtained via interpolation of TRM $m_{DC}$ vs. $T$ data (Fig. 6a) for 2.5 kOe $\leq |H_{DC}| \leq$ 10 kOe and 260 K $\leq T \leq$ 305 K using $m_{DC} = a(T_N-T)^\beta$, where $T_N$ = 300 K, $a$ is a fitting parameter, and $\beta = 0.8$ is the surface critical exponent [6, 15]. The surface moment was fully saturated for the whole $H_{DC}$ range (Fig. 6b), therefore we used an averaged $|m_{DC}(T = 297.5 \text{ K})| = (8.1 \pm 0.2) \times 10^{-11}$ Am². By substituting these values and using $H_{fc}$ = 50 kOe and $E_{fc}$ = 60 kV/cm, we obtained for the Zeeman and ME energies 0.40 ± 0.01 nJ and 0.19 ± 0.02 nJ, respectively. That is, the Zeeman energy was higher under the given conditions, thus preventing the switching via the ME energy from the negative towards the positive ME susceptibility $\alpha$. When replacing $m_S = \sigma S$ and $V = dS$, where $\sigma$, $S$, and $d$ are the surface moment density, film area and film thickness, respectively, then the minimum $E_{fc}$ required to achieve a ME switching is

$$E_{fc} > \mu_0 \sigma / (\alpha d), \tag{1}$$

or $E_{fc} > 130 \pm 16$ kV/cm, if the above parameters are used. This value is higher than $E_{fc}$ = 60 kV/cm that we could apply in our experiments without destroying the sample.

Interestingly, it has already been found that the exchange bias systems made from (0001) $Cr_2O_3$ films (150 nm - 250 nm) require higher field products $E_{fc}H_{fc}$ to achieve AF single domain switching via ME FC [8, 11 - 13]. In those systems the AF single domain type is responsible for the switching of the FM hysteresis bias, characterized via the so-called exchange bias field $H_{EB}$. The microscopic origin is the exchange interaction at the FM-AF interface. Toyoki *et al.* [11, 12] reported, for their exchange bias systems with 150 nm $Cr_2O_3$, switching products $E_{fc}H_{fc} \geq 4212$ kOe·kV/cm and $E_{fc}H_{fc} \geq 6714$ kOe·kV/cm for the textured and twinned $Cr_2O_3$ films, respectively. By recalculating the switching product for 150 nm instead of 500 nm thickness, we obtain $E_{fc}H_{fc} > (4318 \pm 532)$ kOe·kV/cm, that is, our estimate is close to the experimental values if $H_{fc}$ = 10 kOe is chosen. Our results also contradict the hypothesis made in Refs. [11, 12] to explain the increased switching field products, that $Cr_2O_3$ films might have ME susceptibility values much lower than those in the bulk. By another comparison with the exchange bias switching experiments for exchange bias systems with 250 nm $Cr_2O_3$, the field switching products were reported to be 400 kOe·kV/cm and 2000 kOe·kV/cm in Refs. 8 and 13, whereas our estimates yield $E_{fc}H_{fc} > (2591 \pm 319)$ kOe·kV/cm for $d$ = 250 nm and $H_{fc}$ =10 kOe. That is, we have a rather good agreement with the results in Ref. 13. In addition, the exchange bias samples prepared for characterization via the anomalous Hall effect were patterned into areas much smaller than our samples area, cf. microdots with a 20μm diameter or 40μm × 40μm cross point area in Refs. [11, 12] and Ref. 13, respectively. This could have had an additional effect on the $Cr_2O_3$ ME switching characteristics. Typical lateral dimensions of AF domains in bulk (0001) $Cr_2O_3$ are ≥ 1 mm [32].

An additional factor that could affect the results is the possible pinning effects for AF domains, reported, for example, in bulk single crystals of $Cr_2O_3$ [32]. Pinning could result in an energy barrier $W_b$ that prevents the formation of a reversed domain state upon ME FC from above $T_N$. Pinning sites could consist of crystallographic defects in the bulk crystal or film. Adding this energy barrier per film area, $w_b = W_b/S$, to Eq. (1) would modify it to

$$E_{fc} > \mu_0 \sigma / (\alpha d) + w_b/(\alpha d) \times (1/H_{fc}). \qquad (2)$$

Thus, the existence of a switching barrier would further increase the minimal $E_{fc}$ values required for a ME switching and by extension would provide an additional functional dependence on $H_{fc}$. The same functional dependence was found in [11, 12] for exchange bias systems with 150 nm (0001) $Cr_2O_3$. A comparison between twinned and textured $Cr_2O_3$ films in [11,12] yielded different slopes and intersection values for $E_{fc}$ vs. $1/H_{fc}$ dependences. Those differences could be then explained by a distinction between energy barriers and surface magnetization in the $Cr_2O_3$ films of different structural quality, respectively.

**Summary**

The ME effect was measured on 500 nm $Cr_2O_3$ films grown by rf sputtering between two thin film Pt electrodes on $Al_2O_3$ (0001) substrates. The ME response was linear with respect to the applied electric field and similar in magnitude to the response observed for bulk single crystals. The relevant temperatures in the temperature dependence of the ME susceptibility were lower than for thick single crystals of $Cr_2O_3$. We were able to switch the sign of the ME susceptibility by cooling in magnetic fields of opposite directions. Attempts to switch the sign of the ME susceptibility via the so-called ME FC were not successful, presumably as a result of competition between the total Zeeman energy associated with the film surface and the total ME energy associated with the film volume.

**Acknowledgements**

This work was supported in part by a Research Challenge Grant from the WV Higher Education Policy Commission (HEPC.dsr.12.29), WVU Shared Research Facilities, and CSTI ImPACT program through government of Japan.

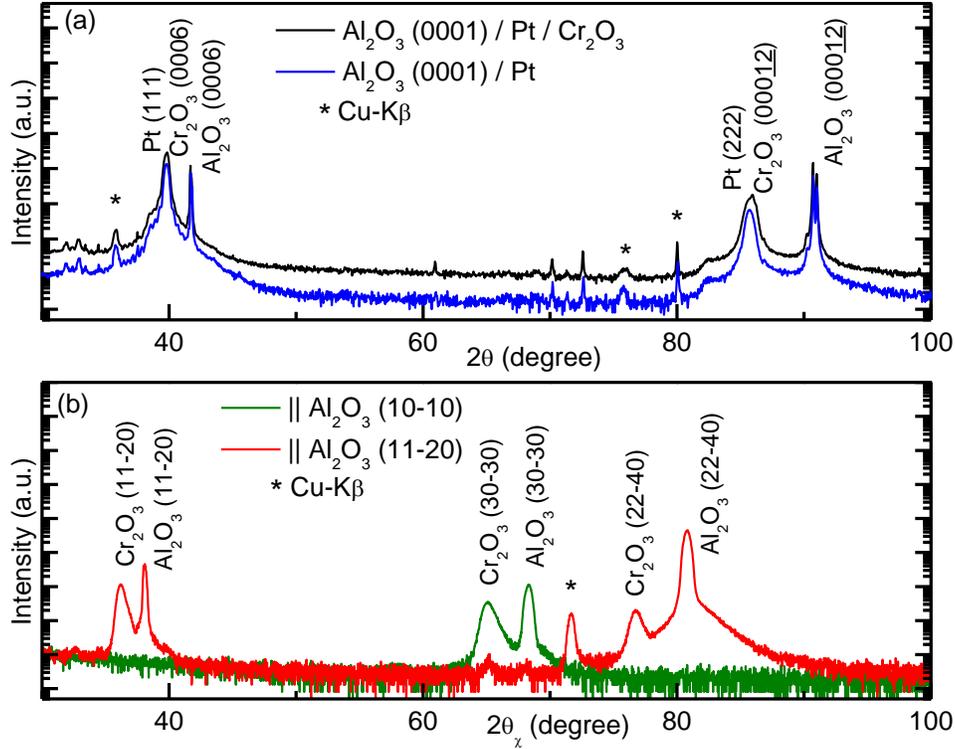

Figure 1. X-ray diffraction structural characterization: (a) XRD out-of-plane $\omega$-$2\theta$ scan from a 500 nm Cr₂O₃ film grown on Al₂O₃ (0001)/ Pt 25 nm substrate (black line). For comparison, XRD pattern from Al₂O₃ (0001) / Pt 25 nm is shown (blue line). Cr₂O₃ (000$l$) peaks overlap with Pt (111) and (222) reflections. (b) XRD $\phi$-$2\theta_\chi$ in-plane scan for Al₂O₃ / Pt 25 / Cr₂O₃ 500 (nm) sample along Al₂O₃ (10-10) [green line] and (11-20) [red line] planes. The corresponding azimuthal ($\phi$) angle was adjusted in order to observe different crystallographic planes.

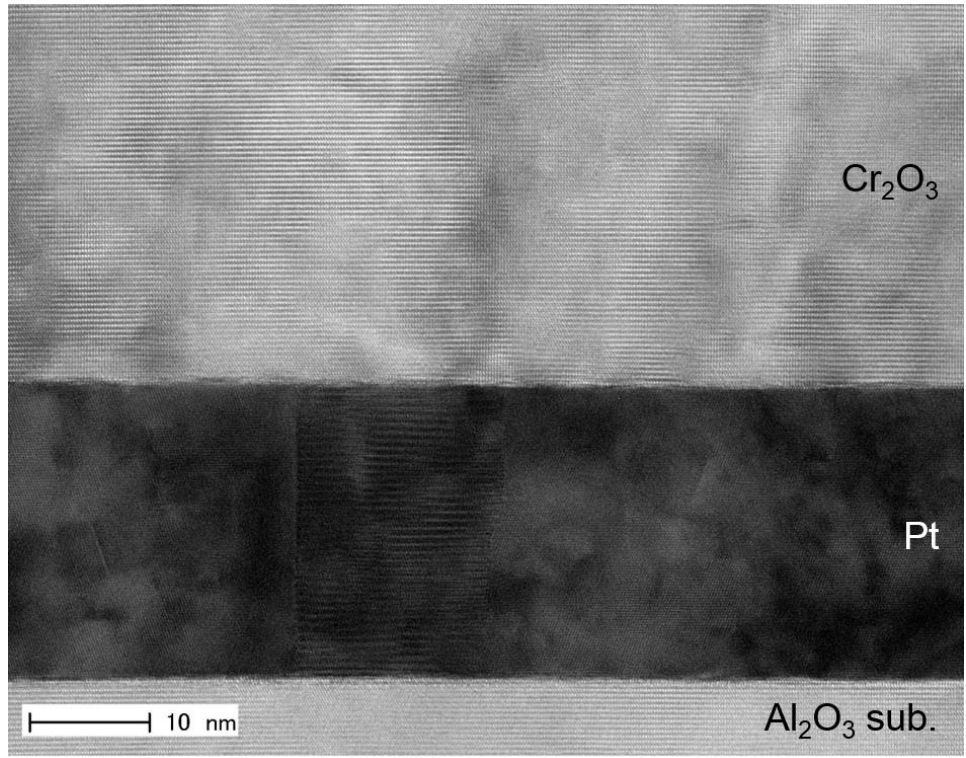

Figure 2. Cross-sectional TEM image on (0001) $Al_2O_3$ / Pt 25 nm / $Cr_2O_3$ 250 nm. The discontinuous pattern on Pt layer indicates its texture.

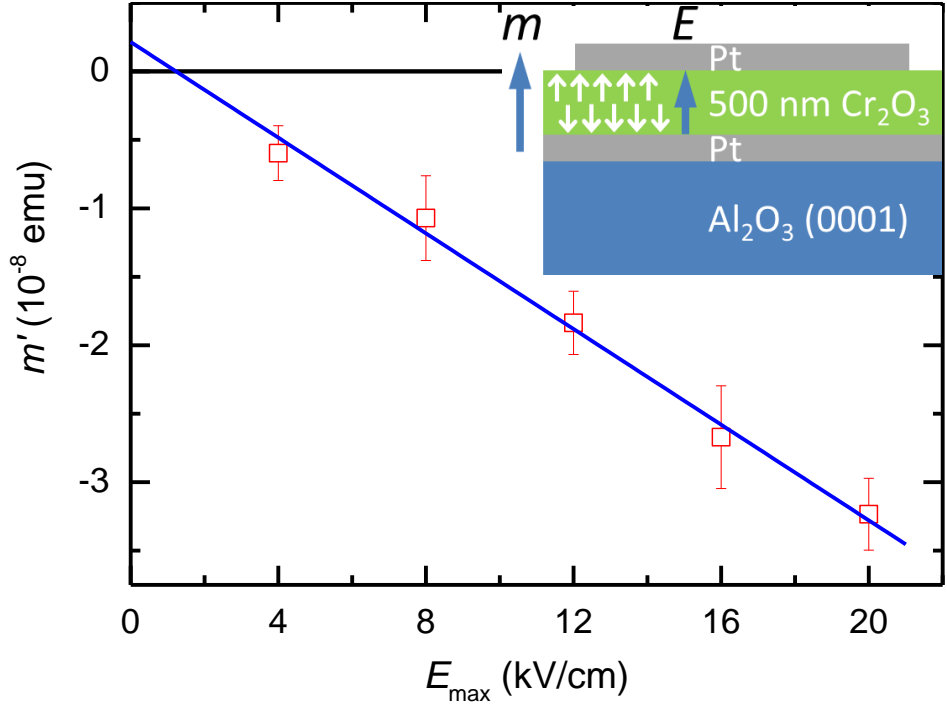

Figure 3. AC magnetic moment amplitude $m'$ vs. AC electric field amplitude $E_{max}$ measured on 500 nm $Cr_2O_3$ film at 250K after cooling from 310K in 10 kOe magnetic field. The blue solid line is the corresponding linear fit. The slope obtained from the fit is proportional to the linear ME susceptibility. Top right inset: illustration of the thin film cross-section structure together with magnetic moment (*m*), electric field (*E*), and $Cr^{3+}$ spin vectors orientations.

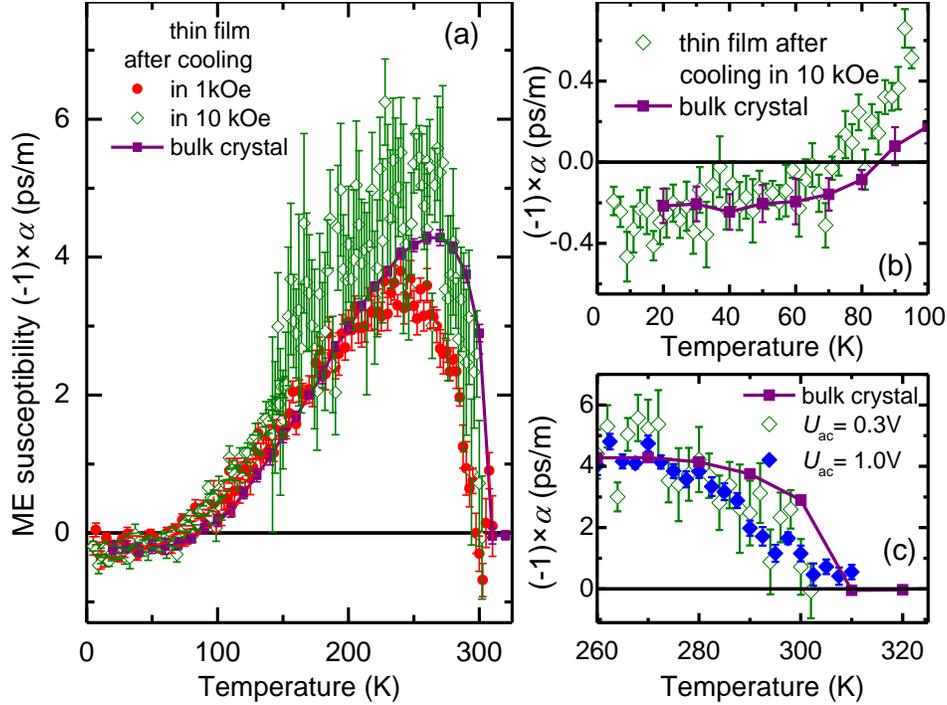

Figure 4. ME susceptibility $\alpha$ vs. temperature. (a) Comparison between a 500 nm $Cr_2O_3$ film and a thick single crystal (0.5 mm) of $Cr_2O_3$. The film was cooled down from 320 K down to 100 K in either 1 kOe and measured using AC electric field amplitudes of 62.4 kV/cm (between 5K and 95K), 31.2 kV/cm (between 97K and 151K) and 20 kV/cm (between 152.5K and 310K); or field cooled in 10 kOe and then measured using AC electric field amplitudes 62.4 kV/cm (between 5K and 95K), 31.2 kV/cm (between 95K and 151K) and 6.24 kV/cm (between 142K and 302K). The thick single crystal was ME cooled from 320 K down to 260 K in 30 kOe and 1 kV/cm. (b) Low temperature data from (a) for film cooled in 10kOe and for bulk single crystal. (c) High temperature data for bulk single crystal in (a) and for film cooled in 10 kOe and then measured either using 6.24 kV/cm (i.e. $U_{ac}$=0.31V) or 20 kV/cm (i.e. $U_{ac}$=1.0V) AC field amplitudes. In all figures, experimental data points for α were multiplied by (-1) for better representation.

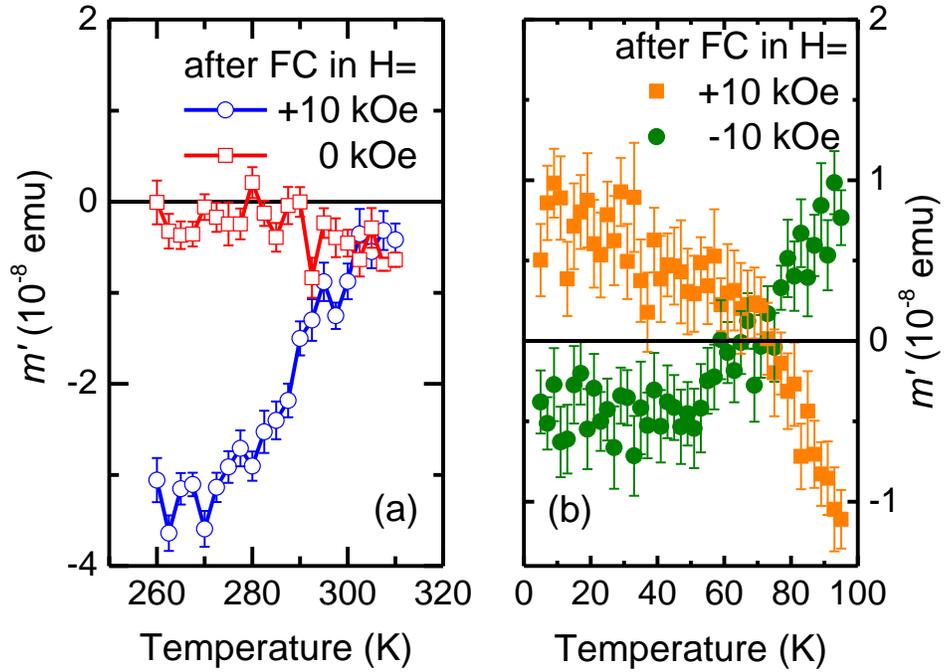

Figure 5. Electrically induced magnetic AC moment amplitude *m'* vs. temperature measured under different FC protocols on a 500 nm $Cr_2O_3$ film. (a) Measurements in the 260 K to 310 K temperature range with an electric field AC amplitude of 20 kV/cm. (b) Measurements in the 5 K to 95 K temperature range with an electric field AC amplitude of 62.4 kV/cm.

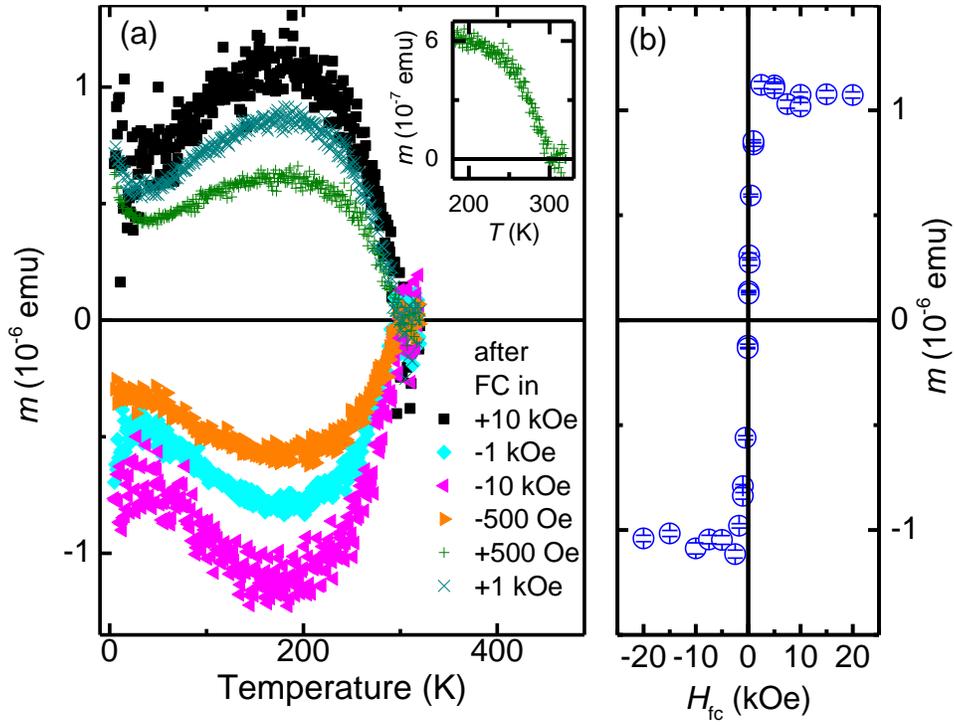

Figure 6. TRM measured after FC from 320 K down to 5 K in different values of $H_{fc}$. (a) TRM magnetic moment $m_{TRM}$ vs. temperature. The fields are listed from top in the same sequences as applied in the experiment. Inset: enhanced view for $H_{fc}$=500 Oe. (b) Magnetic moment $m_{TRM}$ vs. $H_{fc}$ averaged for temperatures between 195 K and 205 K.

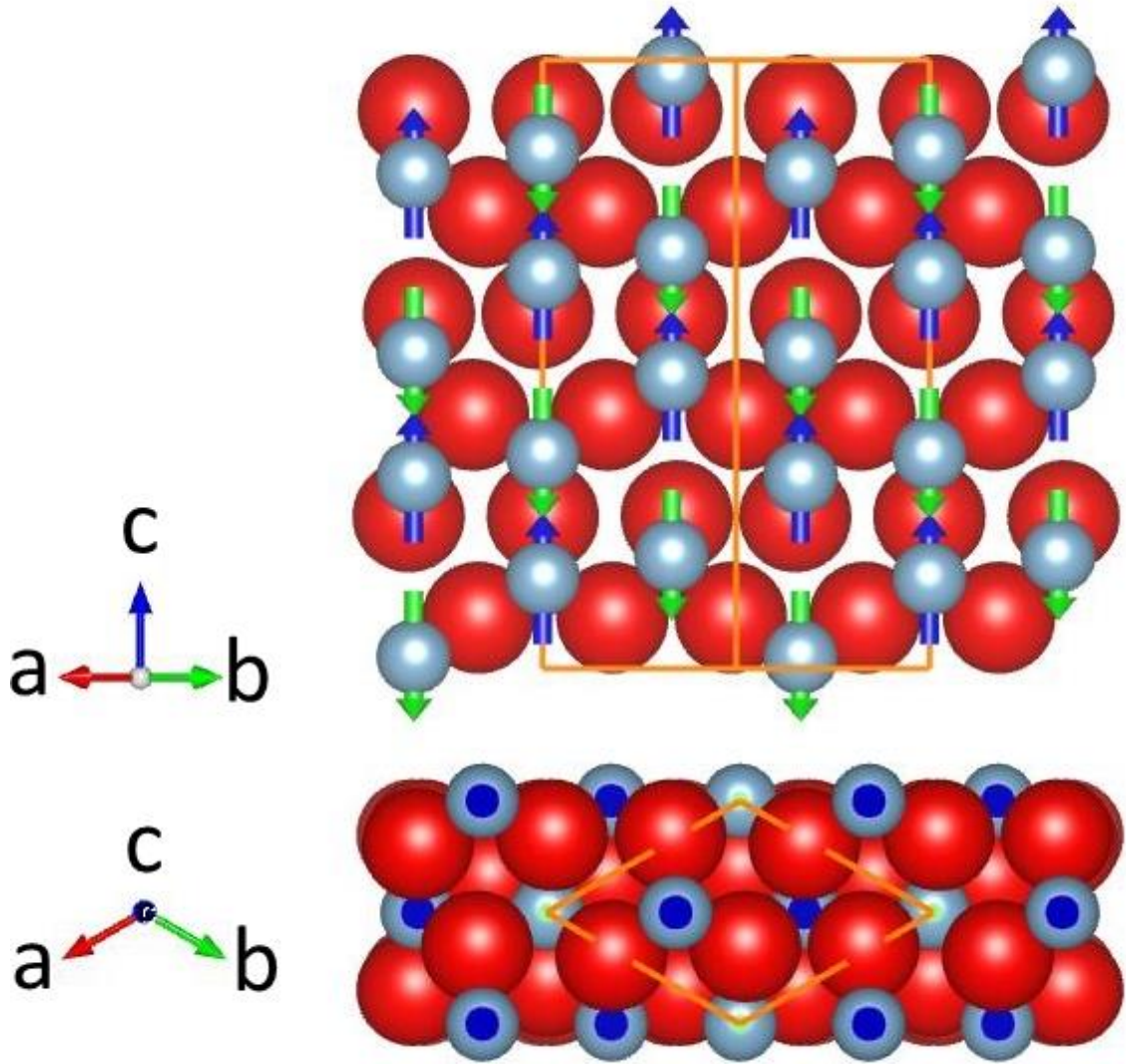

Figure 7. Top: sketch of the (11-20) $Cr_2O_3$ surface. Bottom: sketch of the (0001) $Cr_2O_3$ surface. Large red and Small gray spheres represent $O^{2-}$ and $Cr^{3+}$ ions, respectively. Up- and down arrows denote magnetic $Cr^{3+}$ moments oriented according to the AF spin order in bulk $Cr_2O_3$. Unit cell projections are drawn with thick orange lines. The spin structure shown represents one of two 180° AF domains, labeled type B and corresponding to negative ME susceptibility at $T > 80K$ and to a positive top boundary magnetization along the [0001] axis. A half-cut through the buckled plane of $Cr^{3+}$ ions between two adjacent oxygen layers is taken as a (0001) surface termination.